\documentstyle[aps,harvard,floats,psfig]{revtex}

\newcommand{\beq}{\begin{equation}}
\newcommand{\eeq}{\end{equation}}

\newcommand{\Msun} {\, {\rm M_\odot}} 
 
\newcommand{\kms} {\, {\rm km \, s^{-1}}} 
\newcommand{\kpc} {\, {\rm kpc}} 
\newcommand{\mpc} {\, {\rm Mpc}} 
\newcommand{\Nexp} {N_{\rm exp}}
\newcommand{\Ntil} {\tilde{N}} 
 
\newcommand{\etal} {{\em et al.}}
\newcommand{\that} {\widehat{t}\, } 
\newcommand{\umin} {u_{\rm min}} 
\newcommand{\Amax} {A_{\rm max}} 
\newcommand{\tmax} {t_{\rm max}} 
\newcommand{\vperp} {v_{\perp}} 
\newcommand{\eff} {{\cal E}}
\newcommand{\ten}[1] {\, \times 10^{#1}} 

\newcommand{\simlt} 
     {\mathrel{\hbox{\rlap{\hbox{\lower4pt\hbox{$\sim$}}}\hbox{$<$}}}}
\newcommand{\simgt} 
     {\mathrel{\hbox{\rlap{\hbox{\lower4pt\hbox{$\sim$}}}\hbox{$>$}}}}

\newcommand{\mnras} {Monthly Notices Royal Astron. Soc. } 
\newcommand{\aap} {Astron. Astrophys. } 

\citationstyle{dcu}
\begin{document}

%\special{papersize=8.5in,11in}
\title{Gravitational Microlensing - A Report on the MACHO Project}
\author{Will Sutherland}
\address{Astrophysics, Dept. of Physics, University of Oxford, \\
   Oxford OX1 3RH, U.K. \\}
\maketitle
\begin{abstract}
There is abundant evidence that the mass of the Universe is dominated by 
dark matter of unknown form. 
The MACHO project is one of several teams 
searching for the dark matter around our Galaxy in 
the form of Massive Compact Halo Objects (MACHOs). 
If a compact object passes very close to the line of sight to a 
background star, the gravitational deflection of light causes an apparent
brightening of the star, i.e. a gravitational `microlensing' event.
Such events will be very rare, so millions of stars must be monitored
for many years. 
We describe our search for microlensing using a very large CCD camera 
on the dedicated 1.27m telescope at Mt.~Stromlo, Australia:  
currently some 14 events have been discovered towards the Large
Magellanic Cloud. The lack of short-timescale events excludes
planetary mass MACHOs as a major contributor to the dark matter, 
but the observed long events (durations 1--6 months) 
suggest that a major fraction may be
in fairly massive objects $\sim 0.5 \Msun$. It is currently
difficult but not impossible to explain these events by other
lens populations; we discuss some prospects for clarifying the nature
of the lenses.

\end{abstract}
\vskip 20pt
\centerline{Revised - 4 Jun 98 }
\newpage
\tableofcontents
\newpage
%\pagestyle{myheadings}

%% 0000000000 %% bookmark for start of text
\section{Introduction: Dark Matter}
\label{sec-intro}

This section gives an overview of the evidence for dark matter.  
This is a very large subject so only a brief outline can be given here. 
Further details and references may be found in, e.g., \citeasnoun{popc}.  

There are several strong lines of 
observational evidence for the existence of large quantities
of dark matter in the universe. 
This is often parametrised in units of the critical density by 
$\Omega =  \rho / \rho_{\rm crit}$, where 
$\rho$ is the average density of matter in the universe, and 
$\rho_{\rm crit} = 3 H_0^2 / 8 \pi G 
= 2.78 \ten{11} \, h^2 \Msun \mpc^{-3}$  is the critical density
at which (for zero cosmological constant) 
the Universe is balanced between indefinite 
expansion and eventual recollapse\footnote
{We use standard astrophysical units throughout: 
$G$ is Newton's constant, 
$\Msun \approx 2.0 \ten{30} {\,\rm kg}$ 
is the mass of the Sun, 1 AU $ \approx 1.5 \ten{8} {\, \rm km}$  
is the mean Earth-Sun distance, 1 parsec (pc) $= 648,000 / \pi $ AU, 
3600 arcsec = 1 degree. 
$H_0$ is the Hubble constant,   
parametrised by $h \equiv H_0 / (100 \kms \, {\rm Mpc^{-1}}) 
\sim 0.5 - 1.0$. 
Note that the Sun is $\approx 8 \kpc$
from the center of the Galaxy, the Magellanic Clouds are at 
$\approx 50 \kpc$, and the Andromeda galaxy at $\approx 800 \kpc$. 
}. 
Generally $\Omega_0$ denotes the total matter density, 
and, e.g., $\Omega_B, \Omega_{\rm stars}$ 
denote the fraction of critical density
contributed by baryons, stars, etc.

\subsection{Evidence for Dark Matter}
 
The rotation velocities of spiral galaxies as a function of
galactocentric distance can be accurately measured from the Doppler
effect: at large radii where the stellar surface brightness is falling
exponentially, velocities are obtained for clouds of 
neutral hydrogen using the 21 cm hyperfine line. 
The resulting `rotation curves' are found to be roughly 
flat out to the maximum observed radii $\sim 30 \kpc$, 
which implies an enclosed mass increasing linearly with radius. 
This mass profile is much more extended than the distribution 
of starlight, which typically converges within $\sim 10 \kpc$; 
thus, the galaxies are presumed to be surrounded by extended 
``halos'' of dark matter (e.g. Ashman 1992).

% \subsection{Clusters of Galaxies}

Perhaps the most compelling evidence for dark matter comes from
clusters of galaxies. 
% since there are three independent methods
% for estimating their masses. 
These are structures of $\sim 1 \mpc$ size 
containing $\simgt 100$ galaxies, representing an overdensity of 
$> 1000$ relative to the mean galaxy density. 
They may be assumed to be gravitationally bound since the crossing
times for galaxies to cross the cluster are only $\sim 10\%$ of the
age of the Universe. 
Their masses can be estimated in three 
independent ways:  
\begin{description}
\item[i)] From the virial theorem using the radial velocities of 
individual galaxies as `test particles'. 
% This assumes the clusters are sufficiently old and relaxed for 
% the virial theorem to be satisfied,
% which is reasonably well supported by cosmological simulations. 

\item[ii)] From observations of hot gas at $\sim 10^7 K$ 
contained in the clusters, which is observed in X-rays
via thermal bremsstrahlung. The gas temperature 
is derived from the X-ray spectrum, and the density profile from
the map of the X-ray surface brightness. 
Assuming the gas is pressure-supported against the gravitational
potential leads to a mass profile for the cluster.  

\item[iii)] From gravitational lensing of background objects by the
cluster potential. There are two regimes: the `strong lensing' 
regime at small radii, which leads to arcs and multiple images, and the
`weak lensing' regime at large radii, which causes background galaxies to be 
preferentially stretched in the tangential direction. 
\end{description}

All these methods lead to roughly consistent estimates for cluster masses, 
(e.g.  Carlberg \etal\ 1998, Blandford \& Narayan 1992); 
visible stars contribute only a few percent of the observed mass,
and the hot X-ray gas only $\sim 10-20\%$, so the clusters
must be dominated by dark matter.   

% \subsection{Cosmological Dark Matter} 

On the largest scales, there is further evidence for dark matter: 
`streaming motions' of galaxies (e.g. towards nearby 
 superclusters such as the ``Great Attractor'')
can be compared to maps of the galaxy density from redshift surveys
to yield estimates of $\Omega$ \cite{pecvel}.  
Here the theory is more
straightforward since the density perturbations are still in the linear
regime, but the observations are less secure. 
A similar estimate may be derived by comparing our Galaxy's 
$600 \kms$ motion, measured from the temperature
dipole in the cosmic microwave background (CMB), 
to the dipole in the density of galaxies.  

There are also some useful guidelines from theory.  
Primordial nucleosynthesis successfully explains the abundances
of the light elements $\rm ^{4}He,\, D,\, ^{3}He \ \& \, ^{7}Li$ 
if the density of baryons satisfies $\Omega_B \approx (0.01 - 0.05)
\, (h/0.7)^{-2} .$
This suggests that baryons do not dominate the universe, but 
(depending on the controversial D abundance) this
is probably higher than the density of visible 
matter $\Omega_{\rm vis} \sim 0.01$, 
so allows the dark matter in galactic halos to be mainly baryonic. 

Furthermore, 
it is easier to reconcile the observed 
large-scale structure in the galaxy distribution with 
the smallness of the microwave background anisotropies if the
universe is dominated by non-baryonic dark matter. 
The theory of inflation (postulated to solve the 
horizon and flatness problems) prefers a flat universe with 
$\Omega_0 + \Omega_\Lambda = 1$, where $\Omega_\Lambda$ is the
dimensionless cosmological constant; thus $\Omega_0 = 1$ is the
most `natural' value, which seems to require non-baryonic dark matter. 
The predictions of inflation should be testable in the next decade with  
observations of CMB anisotropy by the MAP and Planck Surveyor
satellites.  

\subsection{Dark Matter Candidates} 

Some `obvious' dark matter candidates 
are excluded by a variety of arguments \cite{carr}: 
hot gas is excluded by limits on 
the Compton distortion of the blackbody CMB spectrum; 
atomic hydrogen is excluded by 21 cm observations;  
ordinary stars are excluded by faint star counts;  
`rocks' are very unlikely since stars do not process hydrogen into
heavy elements very efficiently; 
hydrogen `snowballs' should evaporate or lead to excessive cratering
on the Moon; and black holes more massive than 
$\sim 10^5 \Msun$ would destroy small globular clusters by tidal effects. 
 
Most viable dark matter candidates fall into two broad classes: 
astrophysical size objects called Massive Compact Halo Objects
or MACHOs, and subatomic particles (a subset of which are 
called Weakly Interacting Massive Particles or WIMPs). 

Each of these classes contains various sub-classes: for MACHOs, 
the most obvious possibility is substellar Jupiter-like objects 
of hydrogen and helium less massive than $0.08 \Msun$. 
Below this limit, the central temperature never becomes 
high enough to ignite hydrogen fusion, so the objects just
radiate very weakly in the infrared due to gravitational contraction; 
thus they are usually known as `brown dwarfs'. 
Other MACHO candidates include stellar remnants such as old (and hence, cool)
white dwarfs, neutron stars, and black holes (either primordial
or remnants). 

For the particle candidates, 
they must clearly be weakly-interacting
to have escaped detection, so possibilities include 
the `axion' (hypothesised to solve the strong CP problem\footnote{
The `strong CP problem' is that CP violation in the strong interaction
is very small. Limits on the neutron electric dipole moment
require an arbitrary QCD phase angle to be zero 
within 1 part in $10^9$. This seems unlikely by chance; it is arranged
by the `Peccei-Quinn mechanism' leading to the axion.
}), 
a neutrino (if one or more flavours has a mass $\sim 10$ eV), 
and the popular `neutralino' which is the lightest supersymmetric
particle, thought to be stable. 
(Note that the term `WIMP' is usually reserved for the latter
particle). 
There are active searches in progress for all of these particles
(e.g. \citeasnoun{jkg}), but 
they will not be discussed here. 

\section{Gravitational Microlensing}
\label{sec-ml}

Even if MACHOs comprise most of the Galactic dark matter, 
they will be very hard to detect directly since they would emit 
very little electromagnetic radiation; future infrared
searches may be able to constrain part of the parameter space, 
but not all. 
Thus, it is more promising to detect their gravitational
field, via its influence on the light from background sources.
In addition to the well-known test of General Relativity (GR) 
by light deflection by the Sun, 
gravitational lensing now has many applications in cosmology:  
the first example of a quasar doubly imaged
by an intervening galaxy was discovered by \citeasnoun{wcw}, 
and some 20 such objects are now known.
More recently there have been many discoveries, many 
using the Hubble Space Telescope, of lensing 
of distant galaxies by intervening clusters.
This takes various forms: sometimes 
multiple images are observed, in other cases 
highly distorted `giant arcs' are found, while 
weaker image distortions are found at larger separations as discussed 
in Section~\ref{sec-intro}. An overview of recent gravitational lensing 
observations is given by \citeasnoun{iau173}.   

The principle of lensing by MACHOs in our Galaxy is very similar, 
but we shall see that the relevant angular separation is much 
smaller so the observable consequences are quite different.  

\subsection{Principle of Microlensing} 

If a compact object of mass $M$ at distance $l$ 
lies exactly on the line of sight to a (small) background
source at distance $L$, the light deflection by GR causes
the source to appear as an `Einstein ring' 
(cf Hewitt \etal 1988) with `Einstein radius' $r_E$ 
in the lens plane; 
the (small) light deflection angle is $\alpha  = 4 G M / c^2 r_E $, 
and geometrical optics gives 
$\alpha = r_E / l + r_E / ( L - l) $, 
thus 
\beq
r_E =  \left[ {4 G M L x (1-x) \over c^2} \right]^{1/2} 
\label{eq-re} 
\eeq 
where $x = l/L$ is the
ratio of the lens and source distances; the 
corresponding Einstein angle is $\theta_E \equiv r_E / l$. 
[Note that $r_E \sim \sqrt{r_S L}$, where $r_S$ is the 
Schwarzchild radius of the lens.] 
As we introduce a small misalignment angle $\beta$ between lens and
source, it is clear that two images will be formed 
on opposite sides of the lens, collinear with the lens
and source, at angular positions 
\beq 
\theta_\pm = 0.5 \left( \beta \pm \sqrt{ \beta^2 + 4 \theta_E^2 } \right)
\label{eq-thpm} 
\eeq
from the lens. 
% For a finite-size source of angular radius $\theta_S$, 
% when $\beta > \theta_S$ 
% the Einstein ring splits into two arcs (with opposite image parities), 
% and as $\beta \simgt \theta_E$ the positive-parity image
% tends to its undeflected shape while the 
% negative-parity image rapidly fades away. 

 For a source star at 
$50 \kpc$ and a lens at $10 \kpc$, the Einstein radius is 
$r_E \approx 8 \sqrt{M/\Msun}$ AU, and $\theta_E \sim 10^{-3}$ arcsec. 
This is far below the resolution of 
ground-based telescopes, for which $\theta_{\rm res} \sim 1$ arcsec 
($5 \mu {\rm rad}$) which is set by atmospheric turbulence or `seeing'.
Thus the doubling of the
star's image is not observable, hence the general term `microlensing'. 
However, lensing preserves surface brightness,
thus the apparent flux of the source is magnified by the ratio
of the (sum of the) image areas to the source area. 
A typical star at $50 \kpc$ has an angular radius
$ \theta_S \sim 10^{-6}$ arcsec, and
thus may usually be treated as a point source. 
The fact that 
\beq 
\theta_{\rm res} \gg \theta_E \gg \theta_S
\eeq
explains
much of the simplicity of microlensing; the first inequality
gives the `micro'lensing, while the second inequality gives 
the simple magnification formula of Eq.~\ref{eq-amp} below.  

For a point source and any axisymmetric lens, 
the magnification $A$ of each image is simply
\beq 
A_i = {\theta_i \over \beta_i} \left.{d\theta \over d\beta}\right|_i . 
\eeq
For the point lens, 
the total magnification factor is \cite{refsdal} 
\beq
A = A_{+} + A_{-} = { u^2 + 2 \over u \sqrt{u^2 +4} }
\label{eq-amp}
\eeq
where $u \equiv \beta / \theta_E = b / r_E$ is the 
misalignment in units
of the Einstein radius, and $b$ is the distance of the lens from the
undeflected line-of-sight\footnote{
Two points are noteworthy. Though $A \rightarrow \infty$
as $u \rightarrow 0$, the average magnification integrated over a 
finite source is well-defined. 
Also, though $A > 1$ for all $u$, this does not violate energy conservation
since introducing the lens changes the background metric.}.

Equation~\ref{eq-amp} gives $A \approx u^{-1}$ for $u \simlt 0.5$ and
$A \approx 1 + 2u^{-4}$ for $u \simgt 2$; thus the magnification may be 
very large, but is only appreciable for $u \simlt 2$. 
Of course a constant magnification is not usually measurable, but
since MACHOs must be in motion in the gravitational potential of the
Galaxy, the magnification will be time-dependent due to the changing 
alignment; thus a microlensing `event' will appear as 
a transient brightening with a timescale  
\beq
\that \equiv {2 r_E \over \vperp} \sim 140 \sqrt{M/\Msun} \, {\rm
days};  
\label{eq-that}
\eeq
where $\vperp \sim 200 \kms$ is the transverse velocity of the lens\footnote
{This is estimated from the circular velocity of the Galaxy. 
Though the orbits of dark matter objects are probably not circular, 
their typical speeds must be of this order by the virial theorem.}
relative to the (moving) Earth-source line.  
Assuming constant velocities, 
the apparent `lightcurve' $A(t)$ of the event is simply given by 
Eq.~\ref{eq-amp}  with
\beq
u(t) = \left[ \umin^2 + [2 (t-\tmax)/\that]^2 \right]^{1/2} 
\label{eq-ut}
\eeq
where the minimum misalignment $\umin$ 
(thus maximum magnification) occurs at time $\tmax$. 
This application of microlensing to probe Galactic dark matter was first
suggested by \citeasnoun{pac86}, so 
this form of $A(t)$ is often known as the `Paczynski curve'. 

\begin{figure}[p]
\vspace*{5mm}
%\special{psfile=fig_im.ps angle=-90 voffset=00 hoffset=00
%    vscale=50 hscale=50}
\centerline{\psfig{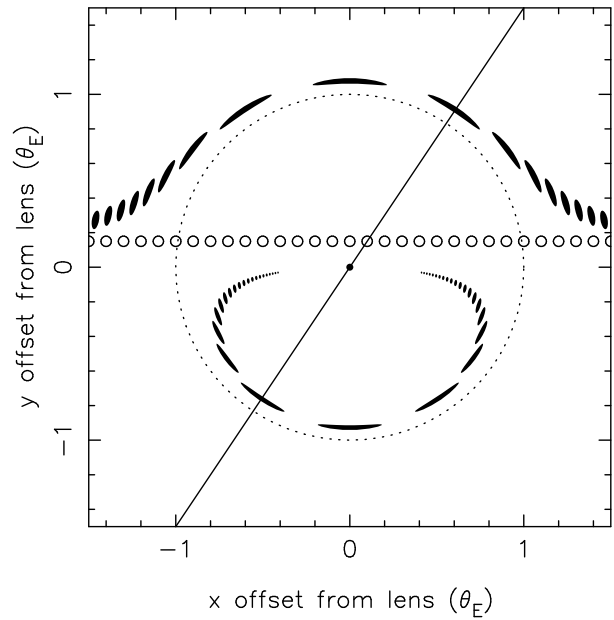}}
\vspace*{5mm}
\caption{A microlensing event seen at `perfect' resolution. 
Axes show angular offsets on the sky of the source from the lens (dot) 
in units of the Einstein angle
$\theta_E$ (defined following Eq.~\ref{eq-re}); 
the dashed circle is the Einstein ring.  
The series of open circles shows the `true' source position 
at successive timesteps, for an event with $\umin = 0.15$.
For each source position, there are two images (solid regions) 
collinear with the lens and source, as indicated by the straight line. } 
\label{fig-ims}
\end{figure}

\begin{figure}[p]
\vspace*{5mm}
%\special{psfile=figevent.ps voffset=-200 hoffset=0 vscale=50 hscale=60}
\centerline{\psfig{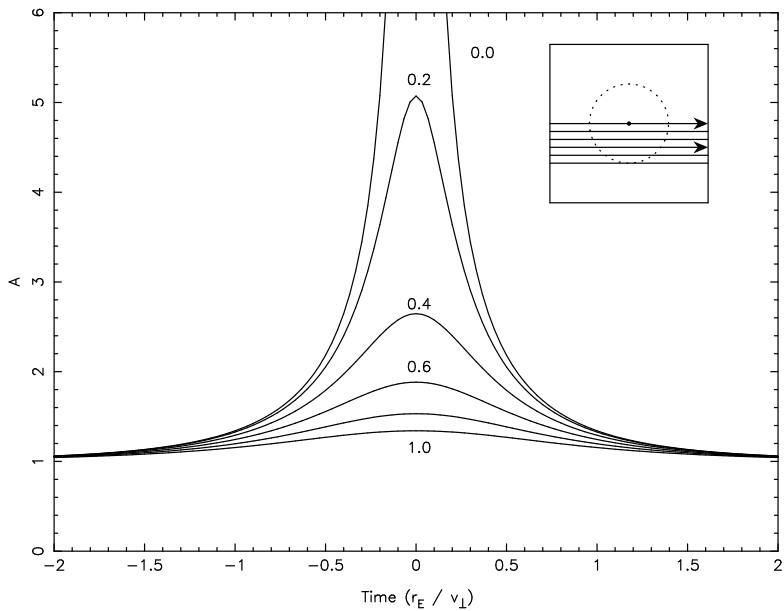}}
\vspace*{5mm}
\caption{ Lightcurves $A(t)$ of microlensing events for 
six values of the impact parameter $\umin = 0.0, 0.2, ... ,1.0 $
as labelled.  
Time is in units of Einstein radius crossing time $r_E / \vperp = \that/2$. 
The inset illustrates the Einstein ring (dashed) and the
source paths relative to the lens (dot) for the 6 curves. } 
\label{fig-paczlc}
\end{figure}

An example of the (unobservable) image behaviour 
for an event with $\umin = 0.15$ is shown in 
Fig.~\ref{fig-ims}, and the lightcurves $A(t)$ for various
values of $\umin$ are illustrated in Fig.~\ref{fig-paczlc}. 
Note that for large magnifications, the events have a distinctive
shape with a sharp central peak and broad wings. 

%FF1 

%FF2

The timescale $\that$ of Eq.~\ref{eq-that} 
is observationally convenient, and the dependence
$\propto \sqrt{M}$ means that a large window in MACHO mass is accessible. 
This window is set at the low mass end by 
the finite size of the source stars, 
which limits the maximum magnification to
$\approx (1 + 4 \theta_E^2 / \theta_S^2)^{1/2}$,  
so lenses of mass $\simlt 10^{-6} \Msun$ cannot 
produce appreciable magnifications. 
At the high mass end the limit $\sim 100 \Msun$
is set both by the patience of the observers and the falling event 
rate (see below). 
This range covers most of the plausible MACHO candidates.  

The microlensing `optical depth' $\tau$ for a given source 
is defined as the mean number of lenses within their own  
Einstein radius of the observer-source line; for $\tau \ll 1$ as here, 
at most one lens gives an appreciable effect, so $\tau$ is just 
the probability that a
random star is microlensed with $u < 1$ or $A > 1.34$ at a given instant; 
thus 
\begin{eqnarray}
\label{eq-tau}
\tau & = & \int_0^L dl \, \int_0^\infty dM 
    \, \overline{n}(l,M) \, \pi r_E^2(l,M) \\
     & = & {4 \pi G L \over c^2} \int_0^L dl \int_0^\infty dM \, 
    M \, \overline{n}(l,M) \, x (1-x) \\
     & = & {4 \pi G L \over c^2} \int_0^L dl \, \rho(l) \, x (1-x)
\end{eqnarray}
where $\overline{n}(l,M) \,dM$ is the number density 
of compact objects at distance $l$ in the mass interval $(M,M+dM)$, 
and $\rho(l)$ is their mass density. 
Since $r_E \propto \sqrt{M}$, the `cross section' of
a lens at given $l$ is $\propto M$. Thus 
$\tau$ depends only on the mass density profile 
$\rho(l)$, and not on the individual lens masses.  
Using the virial theorem, it is straightforward to show that 
$\tau \sim v^2 / c^2 \sim 10^{-6}$
where $v$ is the orbital velocity of the Galaxy; 
more detailed calculations \cite{griest91} 
using a realistic dark matter profile
give
\beq
\label{eq-taulmc} 
\tau_{LMC} \approx 5 \ten{-7}
\eeq
for microlensing of stars in the Large Magellanic Cloud by 
a `standard' dark halo made entirely of MACHOs. 

Note that the rate of microlensing events $\Gamma$ 
does depend on the lens masses via
the durations; clearly the product of the event rate
and the mean duration is proportional to the optical depth. We have 
\beq
\Gamma \langle \that \rangle = {4 \over \pi } \tau \; ,
\label{eq-gam}
\eeq
where the geometrical factor of $4/\pi$ arises because 
$\that$ is defined as the time
for the lens to move by one Einstein diameter, rather 
than the time it spends within the Einstein disk. 
For the LMC this leads to 
\beq
\Gamma \approx 1.6\ten{-6} \sqrt{\Msun/M} \ {\rm events per star per year} . 
\eeq
for a dark halo comprised entirely of MACHOs with mass $M$. 
Thus, high-mass MACHOs would produce very rare long-lasting events, while
low-mass MACHOs would produce relatively more short-duration events.

\subsection{Microlensing Signatures} 

The microlensing optical depth $\tau \simlt 10^{-6}$ is very much
smaller than the fraction of intrinsic variable stars $\simgt 10^{-3}$.
However,  microlensing has many distinct signatures which 
are very different from 
previously known types of variable star: 

\begin{description}
\item[i)] Since the optical depth is so small, only one
microlensing event should be seen in any given star. 

\item[ii)] Gravitational lensing is independent of wavelength, 
 so the star should not change colour during the event. 

\item[iii)] The events should be symmetrical in shape and described 
 by Eqs.~\ref{eq-amp} and~\ref{eq-ut} . These have 3 free parameters, but 
 the event `shape' only depends on $\umin$, while $\that$ and $\tmax$
 just represent a linear transformation of the time axis in
Fig.~\ref{fig-paczlc}.    
\end{description} 
In contrast, most variable stars are periodic or quasi-periodic, 
they are usually asymmetrical in time and show colour changes due to
changing temperatures. 

Also, if many candidate microlensing events can be found, 
 they should satisfy several statistical tests: 

\begin{description}
\item[iv)] Microlensing does not discriminate between types of star, 
 so the events should be distributed across the colour-magnitude
 diagram in proportion to the total number of stars. 

\item[v)] The minimum impact parameter $\umin$ should follow
 a uniform distribution between $0$ and some experimental cutoff $u_T$. 
This translates via Eq.~\ref{eq-amp} 
to a model-independent distribution in peak magnification $\Amax$. 
 
\item[vi)] The peak magnification $\Amax$ and event duration $\that$
 should be uncorrelated. 

\end{description} 

In practice criteria (ii) and (iii) are slightly idealised 
and small deviations may be seen, 
while criteria (iv) -- (vi) are modified by the experimental
detection efficiency, but this can be accounted for as seen later. 

The well-specified shape of microlensing is useful for discriminating
against variable stars, but has the drawback that it limits the
information which can be extracted from each observed event. 
Of the 3 fit parameters $\umin, \that, \tmax$, two give
only the `uninteresting' information of when and how close
the lens approached the line of sight. 
All the desired unknowns 
$M, l, \vperp$ of the lens are folded into the 
single observable $\that$ via Eqs.~\ref{eq-re} and~\ref{eq-that}.
Thus we cannot uniquely determine 
the lens mass or distance from the lightcurve.
If we assume a distribution function in distance
and velocity for the lenses, 
the lens mass may be estimated statistically from
the timescale, but with a large uncertainty; 
roughly a factor of $10^{\pm1}$ in $M$ for a single event. 
Some methods for breaking this degeneracy are discussed later.

\section{Observations}
\label{sec-obs}

The very low optical depth above is the main difficulty of the experiment,
and drives most of the observational requirements. 

Firstly, we require millions of stars at a distance large enough
to give a good path through the dark halo, but small enough that
the stars are not too faint. 
The most suitable targets are the Large and Small Magellanic Clouds
(hereafter LMC and SMC), 
the largest of the Milky Way's many satellite galaxies. Their location
requires a Southern hemisphere observatory. 
The shape of the $A(u)$ function means that it is optimal to monitor
the maximum possible number of stars with $\sim 10\%$ photometric precision, 
rather than fewer stars with high accuracy. 
To get useful statistics, to monitor for long duration 
events and to check the uniqueness of candidates, 
frequent observations over several years are needed,
thus a dedicated telescope is highly desirable;
though a relatively modest (1m-class) telescope is sufficient.   
To check that the observed brightening is 
independent of wavelength, 
it is useful to have simultaneous observations in two different 
wavelength regions.

\subsection{The Telescope \& Cameras} 

These requirements are incorporated in the MACHO experiment 
as follows: we have full-time use of the 1.27-metre telescope 
at Mt.~Stromlo Observatory near Canberra, Australia from 1992-1999; 
it was brought out of mothballs and refurbished
for this project. 
An optical corrector is installed near the prime
focus to give good images over a wide field; this contains
a dichroic beamsplitter to give simultaneous 
images in `blue' and `red' passbands, covering wavelength ranges 
approximately $470 - 630$nm and $630 - 760$nm \cite{macho-tel}. 
Each focal plane is equipped with a very large CCD camera 
\cite{macho-cam} containing
4 Loral CCD chips of $2048^2$ pixels; the field of view
is a square of side $0.7$ degrees.   
Each chip has 2 amplifiers, so the images are read out through
a 16-channel system, 
giving a readout time of 70 sec for 77 MB of data;
the readout noise is $10 e^-$ which is small compared
to photon noise from the night sky. 
The typical exposure times are 300 sec for the LMC, 600 sec for 
the SMC and 150 sec for the Galactic bulge, so around 60-100 images
are obtained per clear night; the stellar detection limit is 
around $21$st visual magnitude, about $10^{-6}$ times fainter 
than the human eye.  All the raw data are archived to Exabyte tape. 
The LMC subtends roughly 6 degrees, so about 80 images are required 
to cover it. The observing strategy has varied so that some fields
were monitored several times per night to be sensitive to
short-duration events, while all fields are observed
at least $\sim 20$ times per year to maximise sensitivity to
long-duration events. 

\subsection{Photometry} 

A typical night's observing produces some 80 images containing
up to 600,000 stars each, so high-speed software is required
to measure their brightnesses. 
We use a special-purpose code called SoDOPHOT,  
customised from the well-known DOPHOT package \cite{dophot}. 
The major modification is the use of `templates' as follows:
for each field
we choose one image from very good sky conditions as a `template image'. 
A `full' reduction is run on this image to produce a list
of detected stars, a `template'.  
Subsequent images are divided into 64 sub-frames called `chunks' 
in each colour, to minimise point-spread function (PSF) 
variation and focal-plane distortions. 
For each chunk, $\sim 30$ bright reference stars 
are located and used to define a PSF, and a coordinate transformation and 
flux scale relative to the template. 
Then, the brightness of all other stars is measured at their
`known' positions using the measured PSF, with neighbouring stars
subtracted from the image. 
Each data point has 6 associated `quality flags' such as
the $\chi^2$ of the PSF fit and the
fraction of the star's image masked due to bad pixels, cosmic rays etc;
these are used later to reject suspect data points. 

This use of `templates' greatly 
speeds up the reductions since for each star there
are only 2 free parameters, the flux and night sky brightness;
a typical image reduction takes around 1 hour on a 
single CPU. 
We now process most of the image data to 
photometry measurements within 12 hours from the time of observation. 
This data is fed to a customised database on 
$\sim 500\,$GB of RAID disks. 

\subsection{Analysis} 

The photometric time-series data is searched for 
variable stars and microlensing
events; over 50,000 variable stars have been found, most of which
are new.  
This very large sample has many benefits for stellar physics, but 
is outside the scope of this article; 
see e.g. \citeasnoun{macho-cook} for an overview. 

For the microlensing search, the lightcurves are first convolved
with filters of various durations, and those showing a peak 
above some threshold are called `level-1' candidates. 
These undergo a full 5-parameter fit to microlensing 
(where the free parameters are 
the 3 microlensing parameters
$\umin,\that,\tmax$ and the 
star's red \& blue baseline fluxes $f_R, f_B$), 
and numerous statistics are computed, including `total significance'
$\Delta \chi^2$, goodness of fit in various windows, number of
significant high points, etc. 
Stars satisfying weak cuts on these are output as `level-1.5' candidates
which can be browsed by eye, and then more stringent cuts are
defined to select a set of final `level-2' microlensing candidates.
The definition of these cuts is necessarily somewhat subjective, 
since little was known in advance about the classes of variable star
which bear some resemblance to microlensing; 
however, in practice
it turns out that after eliminating two sparsely-populated 
regions of the colour-magnitude
diagram (the upper main sequence and the very reddest stars), 
a uniqueness criterion requiring a constant baseline 
and a single high-significance brightnening with $\Amax \simgt 1.5$ 
appears to select a fairly clean set of microlensing events. 

\subsection{Short History of Microlensing} 

While this article focuses on 
results from the MACHO project, 
it is useful to review the progress of microlensing 
including a mention of the other projects; more details
are given in \citeasnoun{pac-rev}. 
At the time of the proposal of \citeasnoun{pac86}, the project
was not technically feasible, but following rapid advances in computers 
and CCD detectors, a seminar by C.~Alcock at 
the Center for Particle Astrophysics, Berkeley in 1990 led 
to the formation of several teams. 
The first teams were MACHO, EROS and OGLE, and
all three announced their first candidate events in late 1993: 
one event towards the LMC by MACHO \cite{macho-nat}, two by EROS
\cite{eros-nat}, and one towards the bulge by OGLE \cite{ogle-1}. 
It soon became clear that the event rate towards the bulge is much
higher than towards the LMC; the world total is now around
150 bulge events and 15 LMC events.  
The majority of these events have
been found by MACHO due to its dedicated telescope and
larger data volume, and are discussed below; 
the OGLE and EROS teams have recently completed
new dedicated telescopes which will increase the event rate. 
Several new groups have
entered the field; DUO observing the bulge \cite{duo2}, 
AGAPE and VATT-Columbia \cite{tomaney} observing the Andromeda galaxy, and 
MOA observing the LMC. Information on these projects may be
found via the MACHO WWW page\footnote{
The MACHO WWW page is 
located at {\tt http://wwwmacho.mcmaster.ca} 
}. 

%%% bbbbbbbbb 

\section{Galactic Bulge Results}
\label{sec-bulge}

It is a lucky coincidence that when the Magellanic Clouds are too low
in the sky to observe, the dense `bulge' of stars 
around the Galactic center is well placed; 
we spend about 1/3 of the observing time on the bulge.
The line of sight to the bulge passes through the disk of our galaxy, 
where the mass density in stars is $\sim 10\times$ 
larger than the halo dark matter density. 
Detection of microlensing does not require that the lens be dark, 
just that it is not much brighter than the source star;
so there must be a significant rate of lensing towards the bulge 
from `known' low-mass stars (actually, these are not directly
observed but must be present assuming a stellar mass function similar
to the solar neighbourhood). 
Thus, the bulge results are only indirectly relevant to the 
dark matter question, but are 
(a) a  useful verification 
that microlensing really occurs and that our experiment can detect it, and 
(b) a probe of galactic structure and the 
low-mass end of the stellar mass function. 

\begin{figure}[!t]
\vspace*{1cm}
%\special{psfile=figevent.ps voffset=-200 hoffset=0 vscale=50 hscale=50}
\centerline{\psfig{width=8cm,file=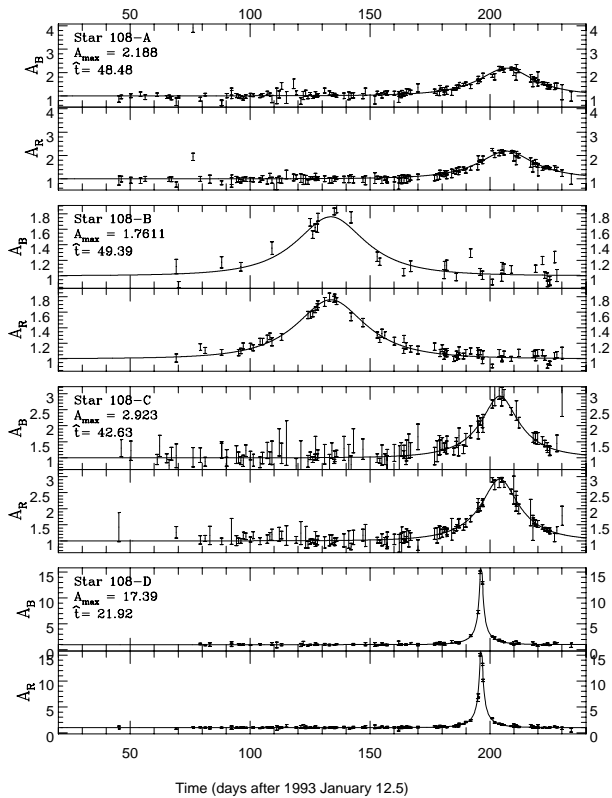}}
\vspace*{5mm}
\caption{
Lightcurves (observed flux vs time) for four representative
microlensing events from the MACHO 1993 season bulge data. 
The data points with $\pm 1\sigma$ error bars show flux in linear units,
normalised to the fitted baseline. 
For each event, the upper/lower panels show the blue/red passbands. 
The smooth curve shows the microlensing fit, simultaneous to both
colours. (From \protect\citeasnoun{macho-blg1}). } 
\label{fig-blg}
\end{figure}

At the time of writing, around 150 candidate microlensing
events have been found towards the bulge; 
43 of these are from the first year's data, for which 
a full statistical analysis has been done \cite{macho-blg1}. 
Some examples of these events are shown in Fig.~\ref{fig-blg}, 
and the $\umin$ distribution 
%% in Fig.~\ref{fig-umin} 
shows excellent consistency with
the predicted uniform distribution, after
correction for the detection efficiency which is higher for small
$\umin$ (large magnification). 

%FF3
%% NB: the [!t]  means Latex actually does what it's bloody told... 

Over $100$ additional events have been discovered by our `Alert System' 
\cite{macho-alert1},  
whereby the photometry is carried out in real time within $12$ hours
of observation. Any star which is not in a pre-defined list of variables, 
shows a $7\sigma$ upward excursion in both colours, and satisfies certain
quality cuts is reported, and the full time series is extracted 
for human inspection. Events judged promising by eye are then 
announced as `alerts', sent to an Email distribution list and placed
on our WWW alert page at {\tt http://darkstar.astro.washington.edu}. 

The main conclusions from the bulge results are as follows: 
\begin{description}
\item[i)] Microlensing is conclusively detected, since many of the events
 are of high quality and the statistical tests are well satisfied. 
 Many events have been observed by multiple sites, and several 
 have been observed spectroscopically in real-time and show 
 no change in the spectrum \cite{bpw} as expected. 
\item[ii)] Estimates of the optical depth towards the bulge 
 range from $2.5 - 3.9 \ten{-6}$
 \cite{ogle-tau,macho-blg1} 
 which is over twice as high as the previous theoretical predictions; 
 this problem is alleviated if the bulge is actually a prolate `bar' 
 seen close to  end-on,  in agreement with recent dynamical evidence. 
\item[iii)] The distribution of event durations is roughly consistent
 with expectation assuming most of the lenses are low-mass stars. 
 There is some evidence for an excess of short timescale events relative
 to predictions \cite{hg-mass}, which may indicate a population of 
 brown dwarfs in the bulge or may be due to blending 
 (Sec.~\ref{sec-finestruc}). 
\item[iv)] Some interesting events are found to show deviations
 from the `standard' Paczynski curve; these are
 discussed in detail in Sec.~\ref{sec-finestruc}.  

\end{description}

%% ssss
\section{Microlensing Fine Structure} 
\label{sec-finestruc} 

The standard `Paczynski curve' assumes 
a single point source, single point lens and uniform relative 
motion between the lens and the observer-source line. 
These are good approximations for most events, but
roughly $10-20\%$ of observed events show significant deviations. 
At first glance one might think that these `deviant' events
cast doubt on the microlensing interpretation, but in practice
many deviations occur for specific subsets of events and have
been predicted theoretically, thus they actually help to
prove microlensing. 
Deviations from the standard shape 
are especially valuable since they can provide additional information
which breaks the intrinsic degeneracy between $M, l$ and
$\vperp$ noted in Section~\ref{sec-ml}.

\subsection{Blending} 

The commonest but least interesting deviation is that caused by
`blending'. Since the fields monitored are chosen to 
have a very high density of stars, 
there is a non-negligible probability that any observed `star'
actually consists of two or more stars within the observational 
seeing disk. Since $\theta_E \ll \theta_{\rm res}$, usually only one
of these will be lensed, so the observed magnification $A_{\rm obs}$ of the
blend will be given by 
\begin{eqnarray}
\label{eq-blend} 
A_{\rm obs} & = & {f_U + A_{\rm true} f_L \over f_U + f_L } \\
A_{\rm obs} -1 & = & (A_{\rm true}-1) {f_L \over f_U + f_L} \nonumber
\end{eqnarray} 
where $f_L, f_U$ are the fluxes of lensed and unlensed stars within 
the seeing disk. 
If the colours of the lensed \& unlensed stars are different, 
the event may appear
chromatic, but there is still a linear relation between passbands. 

Blending can be estimated by extra fit parameters, but unfortunately
there is a near degeneracy in that the lightcurve of a 
blended event appears very similar to that of an unblended event of
lower $\Amax$ and shorter $\that$; post-event Space Telescope images can 
help to break this degeneracy. Blending also affects the detection 
efficiency as discussed later. 

\subsection{Parallax} 
 
The Paczynski curve assumes uniform relative motion between
the lens and the observer-source line. This is clearly inaccurate
at some level due to the influence of the Earth's orbit. 
For events in the observed region of parameter space 
(durations of months), the resulting deviation is small since
a uniform component of the Earth's motion is not separable from
a change in velocity of the lens; 
only the {\em change} in the Earth's velocity $\Delta v_\oplus$ 
over the duration of the event gives a measurable effect. 
The size of the effect scales approximately as 
$\Delta v_\oplus / \tilde{v}$, where $\tilde{v}$ is the
relative velocity of the lens `projected' back to the solar system plane,
$\tilde{v} = \vperp / (1-x)$. 

\begin{figure}[p]
\vspace*{-2cm}  % get rid of the white space at top of fig. 
%\special{psfile=figevent.ps voffset=-200 hoffset=0 vscale=50 hscale=50}
\centerline{\psfig{height=11cm,file=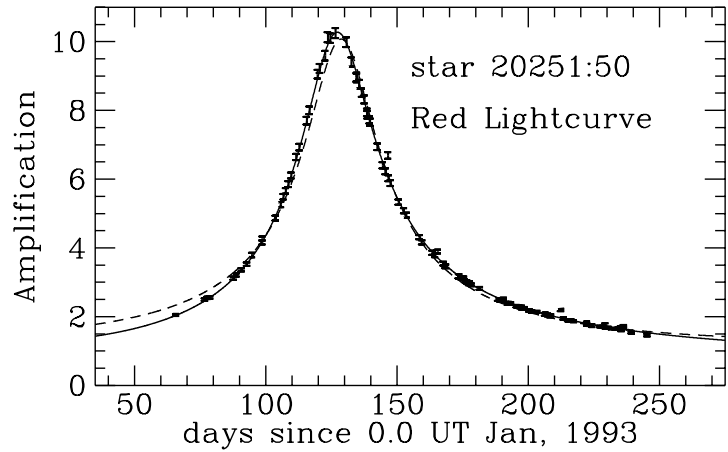}}
\vspace*{5mm}
\caption{The lightcurve of a `parallax' event;  
 data shown as $\pm 1\sigma$ error bars. 
% The upper/lower panels show the red/blue passbands
% (there are fewer points in the blue passband due to a defective area
% on one CCD chip).  
 The dashed curve shows
 the `standard' microlensing fit of Eq.~\protect\ref{eq-ut}, and 
 the solid curve shows the fit including the effect of the Earth's orbit.
 (From \protect\citeasnoun{macho-parallax}). 
} 
\label{fig-parallax}
\end{figure}

This `parallax' effect was predicted by \citeasnoun{gould-parallax},
and first observed by \citeasnoun{macho-parallax}; this was the 
longest event in the MACHO year-1 bulge data, so such a deviation
is not unexpected. 
The light curve of this event is shown in Figure~\ref{fig-parallax}. 
The parallax effect is very useful since it enables the two components
of $\tilde{\bf v}$ to be measured, by reference to the
known parameters of the Earth's orbit. 
Along with the usual $\that$, one
then has two constraints on the three unknowns $M, l, \vperp$ of the lens;  
thus one obtains unique relations for $M, \vperp$ as a function of
assumed lens distance $l$. 
Additional constraints come from likelihood analysis
\cite{macho-parallax} since the probability of observing a given
$\tilde{v}$ is a sensitive function of $l$. 

%FF4 
%% NB - published vers. was non-eps, so used Maryland instead. 

%FF5 
%% This is moved up to appear on same page as Fig.4
\begin{figure}[p]
% \vspace*{1cm}
%\special{psfile=figevent.ps voffset=-200 hoffset=0 vscale=50 hscale=50}
\centerline{\psfig{height=9cm,file=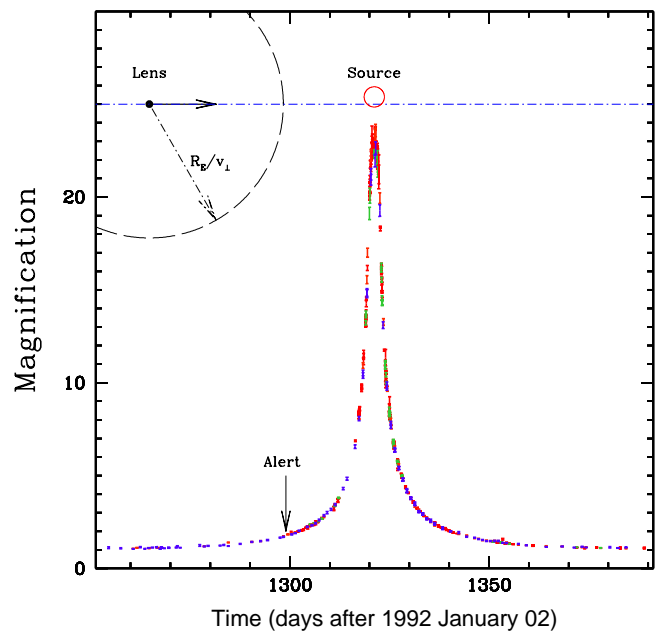}}
\vspace*{5mm}
\caption{ The lightcurve of MACHO Alert 95-BLG-30, which 
 shows finite-source effects. 
 Data points show flux relative
 to the fitted baseline with $\pm 1\sigma$ error bars. 
 The upper part of the figure illustrates the geometry, with 
 the source (small circle) and Einstein ring (dashed circle). 
(From \protect\citeasnoun{macho-9530}).   
} 
\label{fig-9530}
\end{figure}

%A related effect can occur where the source star is a binary (possibly
%with a dark companion); in this case the orbit
%of the two stars can give rise to `ripples' in the observed
%light curve \cite{hg-binsource}, and if
%the orbit can be derived from later radial velocity measurements, 
%useful constraints on lens distance may be found. 

\subsection{Finite Source Effects} 

The usual approximation for $A(u)$ assumes a point source. 
The full expression for a uniform circular disk is given
by \citeasnoun{witt-mao}: in practice this is very close to the
point-source formula except when $\beta \simlt 2 \theta_S$, i.e., the
lens comes close to transiting the disk of the source. 
Since $\theta_E \gg \theta_S$ for solar-mass lenses, this implies
that finite-source effects 
should only be significant for very low-mass lenses or 
near the peak of high-magnification events. 

One very clear example of this effect has been observed,
in MACHO Alert 95-30 \cite{macho-9530}; 
this was an event with $\Amax \sim 25$ involving  
a giant source star. 
The event was detected well before peak, and
subsequent fits predicted a high peak magnification. 
Thus, the possibility of finite-source effects 
was anticipated in advance, and the peak was very frequently 
observed by some 5 observatories around the globe. 

The lightcurve data near the peak is shown in Figure~\ref{fig-9530}, 
which clearly shows the `shoulders' due to the finite-source effect. 
This can be fitted with one additional free parameter, 
the angular size of the star in units of the Einstein angle,  
$u_* \equiv \theta_S / \theta_E$, which is $0.075$ in this example. 
The apparent brightness and spectral type of the 
star give an estimate of $\theta_S$; 
thus we obtain $\theta_E$, which leads to a mass-distance relation
for the lens. 

Such finite source effects can be of considerable astrophysical
interest due to the `differential magnification' effect across
the face of the star; thus, observations as the lens transits the
source can constrain the center-to-limb variations in the star's 
spectrum, which for giant stars is very hard to measure in
any other way. 

\subsection{Binary Lenses}

Undoubtedly the most spectacular deviation from the Paczynski curve
is that arising from a binary lens \cite{sch-weiss,mao-pac}. 
Although $\sim 50\%$ of
all stars reside in binary systems, 
the binary lensing is most dramatic when the projected 
separation $a$ is roughly comparable to the 
Einstein radius; 
%if $a \ll r_E$ the
%binary acts like a single lens of the combined mass, while if
%$a \gg r_E$ it acts like two separate point lenses. 
so the expected frequency of observable binary effects
for stellar lenses is reduced to $\sim 10\%$. 
The first binary lens discovered was OGLE-7 \cite{ogle-7}, 
which was soon confirmed independently 
in the MACHO data \cite{macho-ogle7}. 

\begin{figure}[!ht]
%\special{psfile=figevent.ps voffset=-200 hoffset=0 vscale=50 hscale=50}
\centerline{\psfig{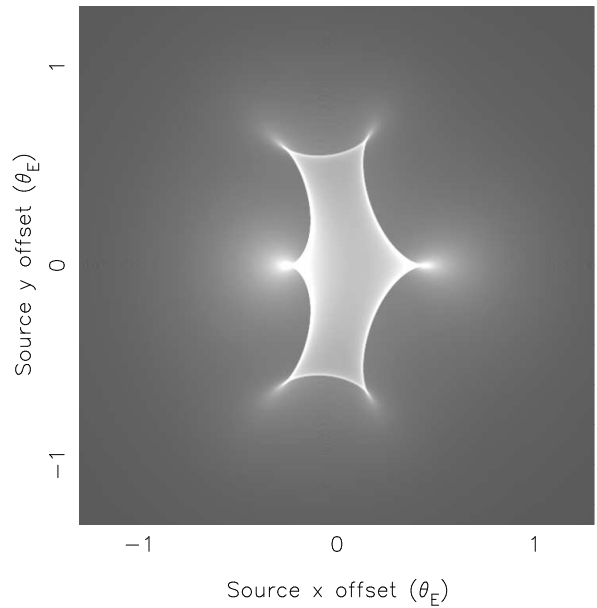}}
\vspace*{5mm}
\caption{Magnification map as a function of source position 
 for a typical binary lens. Axes show source offset from the lens centroid
 in units of Einstein angle $\theta_E$. 
 Grey-scale shows magnification on a log scale, 
 from 1 (black) to 10 (white).  
 The lens consists of two point masses with $2:1$ mass ratio separated
 by $1 \theta_E$ (defined using the total mass of the
 system);  
 the larger mass is at (-1/3, 0) and the smaller at 
 (+2/3, 0). } 
\label{fig-binmag}
\end{figure}

The binary lens is qualitatively different from a single lens
since it contains astigmatism, which breaks the point 
singularity into one or more line singularities, i.e., `caustics'.
These are closed curves in the source plane where the number
of images changes by $\pm 2$. 
Caustics in the source plane map to `critical curves' in the image
plane where the determinant $\vert \partial {\bf \beta} / 
\partial {\bf \theta} \vert = 0$;
hence the magnification is infinite for a point source on a caustic.  
An example magnification map for a binary lens is shown in
Figure~\ref{fig-binmag}; the caustic is the 6-pointed closed curve.
For a point source outside the caustic, there are 3 images; 
as the source crosses the caustic inwards, two new images 
appear `instantaneously' with formally infinite magnification, 
and these fade as the source is well inside the caustic.  
% (For a finite source, one bright new image appears as the source crosses
% the caustic, which splits into two when the source is within the caustic). 

%ff6

Clearly a binary lens can lead to a great diversity of lightcurves,
since there are 3 additional free parameters; the mass ratio, 
the projected separation in units of the Einstein radius, 
and the orientation of the binary relative to the source path. 
However, caustic crossings are common features, which must
occur in inward-outward pairs, so these events are quite
easily recognised.  
% thus, binary microlensing events can be classified into
% `strong' with caustic crossings, `moderate' with multiple peaks
% but no caustic crossing, and `weak' with a single peak but deviations
% from the Paczynski curve. 
Fitting of models to observed binary events is non-trivial due to the
large parameter space and multiple minima \cite{mao-distef}. 
 
To date, about 8 definite binary-lens events have been observed, 
OGLE-7 above, DUO-2 \cite{duo2}, MACHO LMC-9 below and about 5 by  
the  MACHO alert system; there are several more 
probable cases where the data coverage is not quite 
sufficient to be  certain. 

A related effect is the case of lensing by a star with a planetary
system; here the caustics are much smaller, so 
such an event appears like a single-lens event with some 
probability ($\sim 1 - 20\%$) of a short-lived
deviation if one image passes near a planet. This is valuable
since it is sensitive to smaller planet masses \cite{benn-rhie} 
than the recent planet discoveries via radial velocity measurements. 
Thus, to detect planets via microlensing, 
it is desirable to take existing microlensing
alerts and observe them much more frequently ($\sim$ hourly); 
this is being undertaken by two teams, PLANET and GMAN.

\section{LMC Results} 
\label{sec-lmc}

The microlensing searches towards the LMC and SMC are
the most important for the dark matter question, since these lines
of sight pass mainly through the outer Galaxy where the density
is dominated by dark matter; thus, unlike the bulge case, 
the lensing rate for an all-MACHO 
halo is much larger than that from known stars. 

\begin{figure}[!ht]
%\vspace*{2cm}
%\special{psfile=figevent.ps voffset=-200 hoffset=0 vscale=50 hscale=50}
\centerline{\psfig{height=13cm,file=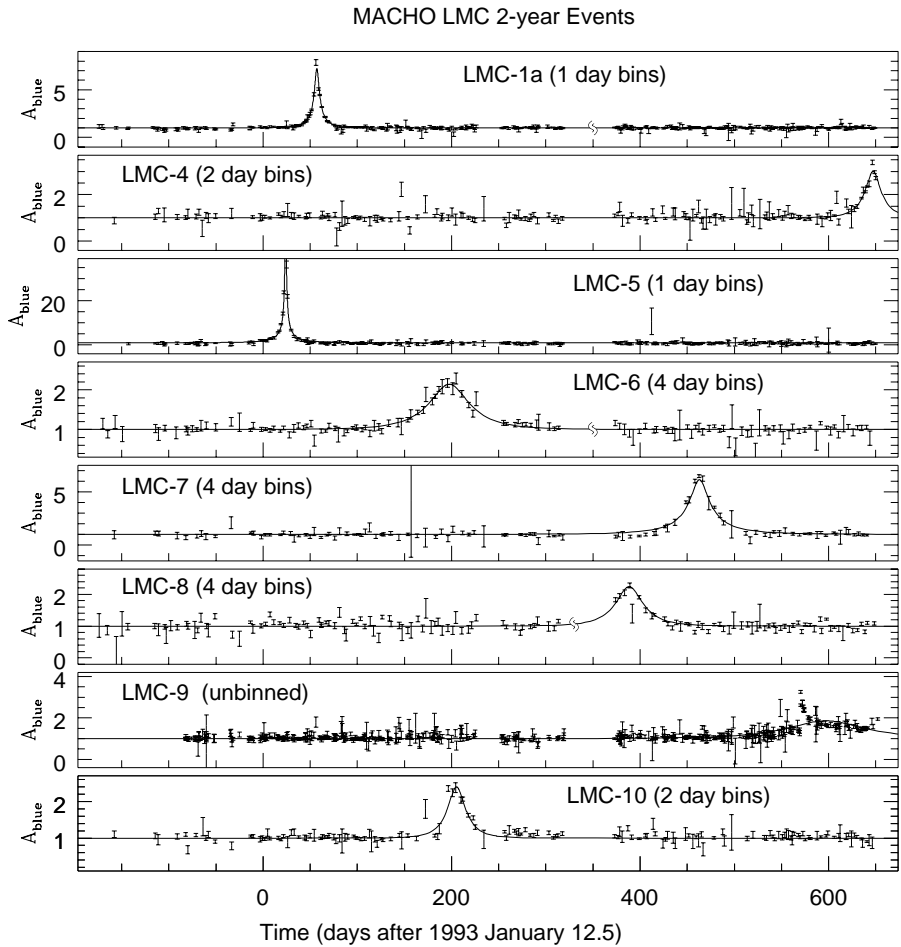}}
\vspace*{5mm}
\caption{Lightcurves (observed flux vs time) for the 
 eight candidate microlensing events from 
 the 2-year Large Magellanic Cloud data. Data points 
 ($\pm 1\sigma$ errors) show flux in linear units, 
 relative to the fit baseline. For clarity, flux data
 have been averaged in time bins of width 1-4 days roughly scaled with the
 event durations (see labels). Only data in one colour
 is shown.  The curves show the single-lens microlensing fit to each event. 
 (From \protect\citeasnoun{macho-lmc2}). 
} 
\label{fig-lmcev}
\end{figure}

At present, we have analysed the first 2.3 years of data\footnote{
Analysis of the 4-year LMC dataset is nearing completion, and yields about 
6 additional events; there are also several further LMC events from the
Alert system. 
These analyses and related efficiencies 
are not yet finalised, but preliminary results indicate an optical
depth similar or marginally lower than the 2-year value. } 
for our `high priority' LMC fields; this sample contains some 8.5 million
stars with 300--800 observations each \cite{macho-lmc2}. 
A set of automated selection criteria applied to this sample
yields 12 objects; 2 of these are redundant detections of
two stars which appear in field overlaps and are independently
analysed, and a further 2 are rejected due to `magnification bias' 
in that they were brighter than normal in our template image and
have subsequently faded below our detection limit (one of
these was almost certainly a supernova in a background galaxy). 
This leaves 8 candidate microlensing events, which are shown in 
Fig.~\ref{fig-lmcev}. (These are numbered 1,4--10 since
two low-quality candidates 2 \& 3 appeared in an earlier paper 
but no longer pass the improved selection criteria). 

%ff7

Here event~1 is the event first announced in \citeasnoun{macho-nat}; 
the star's spectrum shows no abnormalities \cite{dellavalle}. 
Event~4 was discovered by our Alert system, thus has more accurate
follow-up observations (not shown here) 
which support the microlensing interpretation. 
Event~5 occurs in a faint star but has a very high peak magnification;
the magnification appears slightly greater in 
the blue passband than in the red, probably due to blending. 
Event~9 is due to a binary lens; the fit shown in Fig.~\ref{fig-lmcev}
is the single-lens fit used in event detection, thus is not appropriate.
Events~6, 7 and 8 are good-quality candidates, while event~10
appears somewhat asymmetrical and may be a variable star; the inclusion
or exclusion of this event has little effect on the results.  
The distribution of these candidates across the colour-magnitude diagram
and in $\Amax$ is consistent with the microlensing predictions given
in Section~\ref{sec-ml}B. 
Some four of these candidates (1,4,5,9) are of high quality 
and are almost certainly
due to microlensing; 
this suggests that most of the lower quality candidates are
also microlensing, since if {\em only} the high-magnification 
candidates were actually microlensing while the others
were variable stars, the implied distribution of $\Amax$ would
be somewhat improbable. 

Given observed microlensing events, it is natural to ask 
whether the individual lenses can be detected post-hoc, either directly
or by their microlensing other nearby stars. 
Unfortunately, direct detection
is unlikely since the lenses may be $1-6$ orders of
magnitude fainter than the source star, and the relative
angular velocity (``proper motion'' in astronomical jargon)
would be $\sim 0.005$ arcsec/yr; thus even after
a decade they would be only marginally resolved by HST. 
Likewise, it will take centuries for a given lens to reach a 
neighbouring source star, and even then it is improbable
that the alignment would be close enough to lens the second star.

\begin{figure}[!p]
% \vspace*{1cm}
%\special{psfile=figevent.ps voffset=-200 hoffset=0 vscale=50 hscale=50}
\centerline{\psfig{height=8cm,file=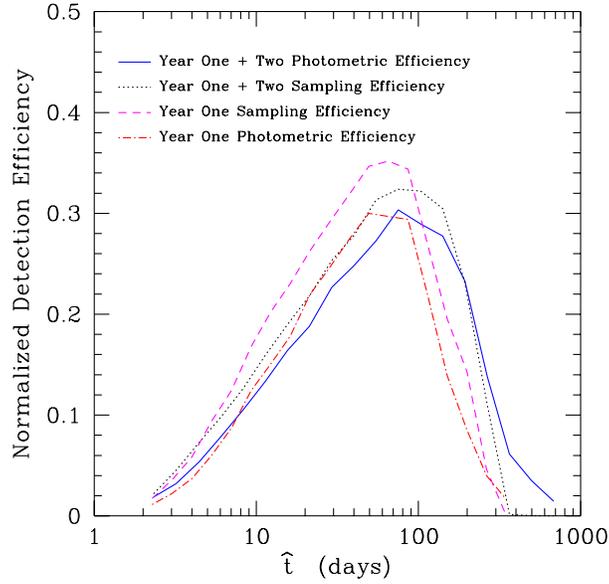}}
\vspace*{5mm}
\caption{
Detection efficiency vs event duration $\eff(\that)$ 
for the LMC 2-year data. The dotted line shows the ``photometric''
efficiency assuming all stars are single, while the solid line
incorporates the effects of blending. 
The efficiency is defined relative
to the theoretical event rate with $\umin < 1$, 
and thus contains a factor
of $0.66$ from the actual cut $\Amax > 1.75$ or $\umin < 0.66$. 
(From \protect\citeasnoun{macho-lmc2}). 
} 
\label{fig-eff}
\end{figure}

To obtain quantitative conclusions, we clearly need to assess
our event detection efficiency. The detection probability for individual
events is a complicated function of $\Amax$, $\that$, the observing
strategy and the brightness of the source star, but all these distributions
except $\that$ are known, and thus can be averaged over 
by a Monte-Carlo simulation. There are two levels of detail here:
firstly a `sampling efficiency' which assumes all stars are single sources
and adds events at the time-series level; and secondly a `blending
efficiency' which adds artificial stars into the raw image data. 
The latter is more realistic, but typically only differs by $\sim 15\%$
from the sampling efficiency, since there is an approximate
cancellation between the increased number of stars in a blend
and the reduction in observed magnification. 
The event detection efficiency $\eff(\that)$ 
is shown in Fig.~\ref{fig-eff}, 
and shows a broad peak between $\that \sim 30 - 300$ days. 

\begin{figure}[!p]
%\vspace*{5mm}
%\special{psfile=figevent.ps voffset=-200 hoffset=0 vscale=50 hscale=50}
\centerline{\psfig{height=11cm,file=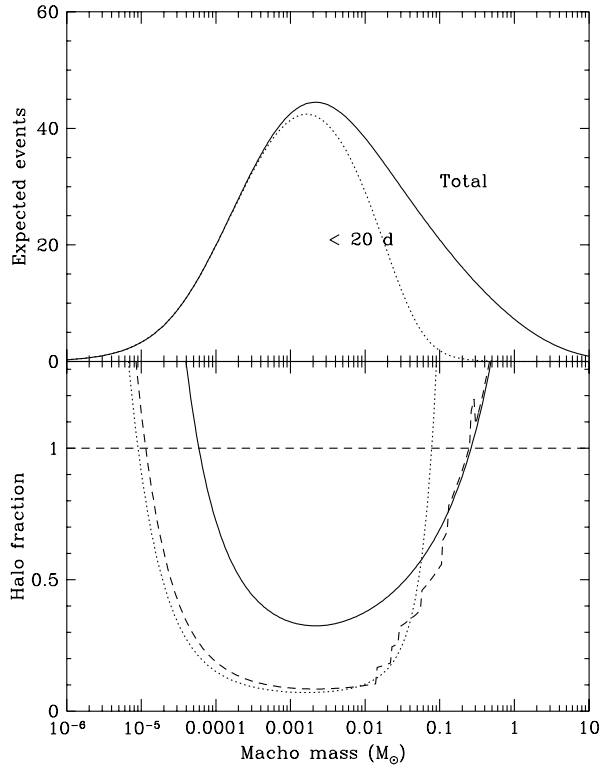}}
% \vspace*{5mm}
\caption{Upper panel: the expected number of microlensing
events vs MACHO mass $\Ntil(M)$, including detection efficiency,  
for an all-MACHO dark halo with unique MACHO mass. 
The solid line shows total number of events, the dotted line
shows events with duration $\that < 20$ days. 
Lower panel: resulting limits on the fraction of the halo 
contributed by MACHOs of given mass; regions above the curves
are excluded at $95\%$ CL. 
The solid line is derived from 8 events total, the dashed line
from no observed events with $\that < 20$ days. 
(From \protect\citeasnoun{macho-lmc2}). 
} 
\label{fig-nexp}
\end{figure}

Given a model for the halo density profile and velocity dispersion, 
it is straightforward to predict the event rate and distribution of
timescales $\that$ for arbitrary lens mass; combined with 
the efficiency function, this gives an expected number of events $\Nexp$. 
In practice it is convenient to 
show the expected number of events $\Ntil(M)$ 
for an all-MACHO halo in which all MACHOs have a unique mass $M$,
which is shown in Fig.~\ref{fig-nexp}. 
Note that $\Ntil(M)$ peaks at $\Ntil > 40$ for
$M \sim 0.001 - 0.01 \Msun$; 
the reason is that for $M \simgt 0.1 \Msun$, 
most events are longer than 30 days where the efficiency curve is
quite flat, while the event rate is falling $\propto M^{-1/2}$. 
For small masses $M  \simlt 10^{-3} \Msun$, 
the theoretical event rate is large but most events are shorter
than 3~days where the efficiency is very low.  The product of these
two factors gives the peak in $\Ntil(M)$. 

%ff8

%ff9 

\begin{figure}[!h]
\vspace*{5mm}
%\special{psfile=figevent.ps voffset=-200 hoffset=0 vscale=50 hscale=50}
\centerline{\psfig{height=8cm,file=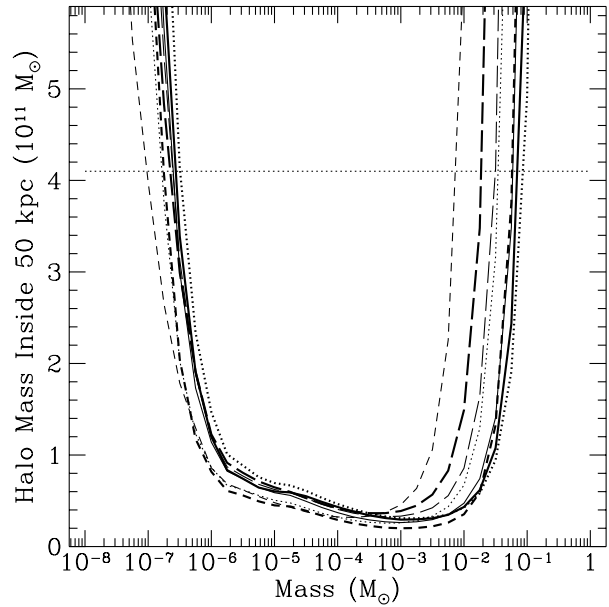}}
\vspace*{5mm}
\caption{
Upper limits on low-mass MACHOs based on the non-detection
of very short duration `spike' events, and 
no events with $\that < 20$ days in the standard search. 
The lines show the
$95\%$ c.l. upper limit on the total mass of MACHOs within 
$50 \kpc$ of the galactic center, as a function of
the assumed MACHO mass, for 8 different models of the dark halo. 
(From \protect\citeasnoun{macho-spike}). 
} 
\label{fig-spike}
\end{figure}

There are several complementary ways to analyse a 
set of detected microlensing events: 
firstly, one may obtain robust upper limits for given $M$  
by simply excluding models which predict too many detected events, 
shown in the lower panel of Figure~9. 
For substellar MACHOs we may obtain stronger limits
since most events should have short duration ($\that < 20$ days), 
but we have no candidate event shorter than this. 
The dotted lines in Fig.~\ref{fig-nexp} show the expected
number of events with $\that < 20$ days, and the resulting limit: 
objects in the mass range $10^{-4} - 0.01 \Msun$ comprise
$< 20\%$ of a standard dark halo, and this applies
whatever the shape of the mass function. 
We have extended this constraint to smaller masses using a separate
search for very short duration events, called a `spike' analysis;
since many fields were observed twice per night, 
we can search for sets of four data points (two red-blue pairs)
from a single night which show a significant brightening in 
all four points. 
There is insufficient data here to obtain a microlensing fit, 
so if events were found we could not claim a detection; 
but using suitable selection criteria, no such events are found, 
which allows us to extend the excluded region down to around
$10^{-6} \Msun$, as shown in Fig.~\ref{fig-spike} \cite{macho-spike}. 
Similar (nearly independent) limits have been obtained by EROS
\cite{eros-ccd}. 

%ff10

Secondly, one may obtain an unbiased estimate of the
optical depth by 
\beq
\label{eq-tauest} 
\tau_{\rm est} = {\pi \over 4 E} \sum_i {\that_i \over \eff(\that_i) }
\eeq
where $\that_i$ is the timescale of the $i$th event and
$E$ is the `exposure' in star-years. 
For the LMC 2-year sample this gives 
$\tau_{\rm LMC} \approx 2.9^{+1.4}_{-0.9} \ten{-7} $. 
This quantity has the virtue of being independent of assumptions about 
lens masses and velocity distributions; 
the drawback is that it
is subject to non-Poisson statistics, and also does not account
for contributions from timescales outside
the window of sensitivity where $\eff \rightarrow 0$. 
Thus, we caution that it is not valid  
to convert $\tau_{\rm est}$ into
a constraint on the total abundance of MACHOs 
without specifying an associated mass interval; 
unfortunately this 
over-general statement sometimes appears in the literature. 

\begin{figure}[!h]
%\vspace*{1cm}
%\special{psfile=figevent.ps voffset=-200 hoffset=0 vscale=50 hscale=50}
\centerline{\psfig{height=10cm,file=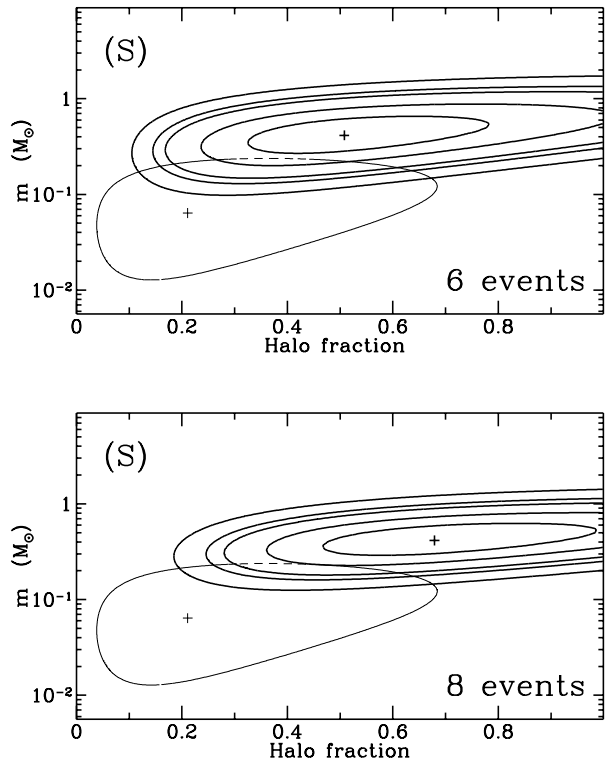}}
\vspace*{5mm}
\caption{
Likelihood contours for MACHO mass $M$ and the fraction of the halo
comprised of MACHOs $f$, for a standard model halo 
with a $\delta-$function MACHO mass distribution. 
The lower panel assumes all 8 events are due to halo lenses, 
the upper panel assumes 6 such events. 
 (From \protect\citeasnoun{macho-lmc2}). 
} 
\label{fig-like}
\end{figure}

The most dramatic result here is that the $\tau_{\rm est}$ above
is not much smaller than the value in Eqn.~\ref{eq-taulmc}, 
and both $\tau_{\rm est}$ and the 
number of 8 observed events substantially exceeds expectations
from lensing by `known' low-mass stars.  
Lensing by stars
is estimated to contribute $\tau_{\rm stars} \approx 0.5 \ten{-7}$ or 
around $1.1$ event to the above sample; hence there appears
to be a very significant excess. 
Thus, one may treat the events as a detection of halo dark matter
and use a maximum likelihood model to estimate the most
probable MACHO mass $M$ and MACHO fraction of the halo $f$. 
Assuming a unique mass for all MACHOs, 
the result is shown in Fig.~\ref{fig-like}, and is 
$M = 0.5^{+0.3}_{-0.2} \Msun$, $f = 0.5^{+0.3}_{-0.2}$. 
A common source of confusion here is that that the 95\% c.l. excluded
region in Fig.~\ref{fig-nexp} is {\sl not} simply the complement
of the $95\%$ allowed region in Fig.~\ref{fig-like}. 
The reason is that the unknown MACHO fraction of the halo 
is really a function $\psi(M)$ (with 
$f = \int \psi(M) dM$),  
not a point in 2-parameter space $f, M$. 
The likelihood analysis depends on the shape of $\psi(M)$ through
the event timescales; this 
is unknown {\em a priori} and  
we must adopt some simple parametrisation, e.g. a $\delta$-function
or a truncated power law.   
If we assume a different shape for the mass function, the `allowed region' 
in the likelihood analysis may change, but the excluded region in
Fig.~\ref{fig-nexp} does not. Thus one needs both analyses to 
extract the full information from the data. 

%ff11

\section{Discussion}
\label{sec-disc}

The interpretation of the LMC results is currently unclear. 
One conclusion is robust, that MACHOs in the planetary mass range
$10^{-6}$ to $0.05 \Msun$ do not contribute a substantial fraction
of the Galactic dark halo. 
Regarding the detected events,  
although the Poisson uncertainties are substantial, 
this is not the dominant uncertainty: the critical question
is, to what population of objects do most of the lenses belong ? 
As noted above, the 8 events in the 2-year data 
are substantially in excess of the 
predicted `background' of $\approx 1.1$ event 
arising from known stellar populations,
which suggests that MACHOs in a dark halo are a natural explanation.

However, there are some astrophysical difficulties with this
interpretation, mainly arising from 
the estimated mass $\sim 0.5 \Msun$ for the lenses.  
These cannot be hydrogen-burning stars in the halo since
such objects are limited to $< 3\%$ of the halo mass by 
deep star counts \cite{gbf}. 
Modifying the halo model to slow down 
the lens velocities can reduce the implied lens mass somewhat, but probably
not below the substellar limit $0.08 \Msun$. 
Old white dwarfs have about the right mass and 
can evade the direct-detection constraints, but
it is difficult to form them with high efficiency, and there may be
problems with overproduction of metals and 
overproduction of light at high redshifts from the luminous
stars which were the progenitors of the white dwarfs \cite{char-silk}. 
Primordial black holes are a viable possibility, though one has
to appeal to a coincidence to have them in a stellar mass range. 

Due to these difficulties of getting MACHOs in the inferred mass range 
without violating other constraints, 
there have been a number of suggestions for explaining the
LMC events without recourse to a dark population:  
most of these suggestions construct some non-standard distribution of 
`ordinary' stars along the LMC line of sight, to increase 
the stellar lensing rate.  
Fairly general arguments from faint star counts can limit
the contribution from our Milky Way disk \cite{gbf}. 
The contribution from LMC `self-lensing', i.e.
stars in the front of the LMC lensing those in the back, is
less well constrained: 
\citeasnoun{sahu} estimated an optical depth $\sim 0.5 \ten{-7}$ from
this, but this has been disputed by \citeasnoun{gould-self}
who obtains an upper limit of $0.25 \ten{-7}$  using dynamical constraints. 
This can be tested given more data via the distribution of events
on the sky, since lenses in the LMC itself 
should be concentrated towards the center of the LMC. 
Other suggested lens populations include 
an unknown dwarf galaxy roughly half-way to the LMC 
\cite{zhao-dwarf}, a 
tidal tail of stars stripped from the LMC \cite{zhao-tail}, 
or a strongly flared and warped disk of our Galaxy \cite{evans-flare}. 
All of these proposals appear somewhat contrived but can be tested 
observationally in the near future. 

However, the most decisive test for the lens population  
is to break the degeneracy in $M,l,\vperp$ 
and thus estimate the distance to at least a subsample of the lenses. 
This can be done for `non-standard' lensing events 
using the methods outlined in Sec.~\ref{sec-finestruc}, but the
percentage of events showing such deviations is expected to be small, 
especially for halo lenses. 
The ideal method for measuring lens locations is to 
discover lensing events in real time, as already happens, 
and then make extra observations from a small satellite at $\simgt 0.3$ AU
from the Earth \cite{gould-2sat}.  
Since this distance is comparable to the Einstein radius of the lens, the
event lightcurve will appear substantially different from the
satellite and the Earth, e.g. the times of peak brightness are expected to
differ by $\sim 1$ week. 
As with the parallax effect of
Section~\ref{sec-finestruc}, this yields a measurement of 
the `projected' velocity of the lens across the Solar system 
$\vperp / (1-x)$. This can unambiguously
determine whether the lens belongs to our disk, the halo or the LMC. 
Another possibility is that planned space astrometric missions 
such as SIM \cite{pac-sim} and GAIA should be able 
to measure the deflection of
the light centroid $\sim 10^{-4}$ arcsec \cite{walker} 
during a microlensing event,
which gives a measurement of the angular Einstein radius $\theta_E$
and thus the lens angular velocity $\vperp /l$. 
Measuring both parallax and $\theta_E$ for the
same event gives a complete solution for $M$, $l$ and $\vperp$.

The prospects for microlensing appear bright: the MACHO project will 
continue to observe until at least 1999, and the EROS-2 and OGLE-2
telescopes have recently come on line, which should significantly
increase the event rate, reducing the statistical
uncertainties and enabling useful tests for the distribution 
of lenses across the LMC. 
Microlensing searches in new directions are also starting to
produce results, including the first event towards the SMC 
\cite{macho-smc1,eros-smc1} and candidate events towards M31 
\cite{tomaney}. 
The suggestions for stellar lensing populations can be tested 
observationally, though if these populations are not found 
we may need to wait a few years for one of the space-based missions
to finally decide whether the lenses are in the dark halo.

\acknowledgments
This paper is dedicated to the memory of Alex Rodgers, who died 
prematurely on 10 October 1997; 
this project would have been impossible without his vision \& 
strong support as Director of Mt.~Stromlo Observatory. 
I am very grateful to all my colleagues in the MACHO collaboration, 
Charles Alcock, Robyn Allsman, David Alves, 
Tim Axelrod, Andrew Becker, Dave Bennett, Kem Cook, Ken Freeman, 
Kim Griest, Jerry Guern, Matt Lehner, Stuart Marshall, 
Bruce Peterson, Mark Pratt, Peter Quinn, Chris Stubbs 
and Doug Welch for a very exciting collaboration and numerous
lively discussions. 
I acknowledge support from the PPARC.

%%% RRRRRRRRR 

\newpage
\clearpage

%%% FFFFFFFFF

\end{document}